\documentclass[aps,prl,twocolumn,showpacs,groupedaddress]{revtex4}
\usepackage{graphics}
\usepackage{graphicx}

\begin{document}


\title{Breaking of Energy Conservation Law: Creating and Destroying of
Energy by Subwavelength Nanosystems}


\author{S.V. Kukhlevsky}
\affiliation{Department of Physics, University of P\'ecs,
Ifj\'us\'ag u.\ 6, H-7624 P\'ecs, Hungary}


\begin{abstract}
The {\em extra energy}, {\em negative energy} and {\em
annihilation of energy} by the subwavelength
conservative systems that have a wave nature of light or matter
(quantum) objects are predicted. The creating and destroying of energy break
the energy conservation law in any subwavelength physical system.
The paradoxical phenomenon is demonstrated in the context of extraordinary transmission of light
and matter through subwavelength apertures [T.W. Ebbesen et al.,
Nature (London) 391, 667 (1998) and E. Moreno et al., Phys. Rev.
Lett. 95, 170406 (2005)].

\end{abstract}

\pacs{42.25.Bs, 42.25.Fx, 42.79.Ag, 42.79.Dj}

\maketitle

%
\maketitle
%
%
The energy conservation law is the most important of
conservation laws in physics. Conservation of energy states that the
total amount of energy in a isolated system remains constant.
In other words, energy can be converted from one form to another,
but it cannot be created or destroyed. The energy
conservation law is a mathematical consequence of the shift symmetry
of time; energy conservation is implied by the empirical fact that
physical laws remain the same over time. The energy conservation affects all physical
phenomena without exceptions, for an example, the
recently discovered extraordinary (enhanced) transmission of light through
subwavelength apertures in a metal screen \cite{Ebb,Schr,Sobn,Port,Taka,Lala,Barb,Leze}.
The transmission enhancement is a process that can
include the resonant excitation of surface plasmons
\cite{Schr,Sobn,Port}, Fabry-Perot-like intraslit modes
\cite{Taka,Lala,Barb}, and evanescent electromagnetic waves
at the metal surface \cite{Leze}. In the case of thin screens
whose thickness are too small to support the intraslit resonance,
the extraordinary transmission of light or matter (electrons) is
caused by the resonant excitation of surface waves \cite{Schr,Sobn,Port,More}.
At the resonant conditions, the system redistributes the
electromagnetic energy around the screen, such that more energy
is effectively transmitted compared to the energy
impinging on the slit opening. The total energy of the system
is conserved under the energy redistribution. In the present paper,
we predict the {\em extra energy}, {\em negative energy} and 
{\em annihilation of energy} by an ensemble of light or matter beams produced
by an array of subwavelength apertures. The creating and destroying of
energy break the energy conservation law in any subwavelength physical 
system. The phenomenon, in particular, is associated with the extraordinary 
transmission without assistance of the surface waves.

Let us first investigate the transmission of light through a
subwavelength structure, namely an array of parallel subwavelength slits.
The array of $M$ independent
slits of width $2a$ and period $\Lambda$ in a metallic screen of thickness
$b\ll\lambda$ is considered. The metal is assumed to be a perfect conductor.
The screen placed in vacuum is illuminated by a
normally incident TM-polarized wave with wavelength
$\lambda=2{\pi}c/\omega=2\pi/k$. The magnetic field of the wave
${\vec{H}}(x,y,z,t)=U(x){\exp}(-i(kz+\omega{t})){\vec{e}}_y$ is
assumed to be time harmonic and constant in the $y$ direction.
The energy balance, which determines the transmission coefficient
of the slit array, is derived by calculating the power of light beams
in the far-field diffraction zone. The EM beams produced by each of the
independent slits are computed by using the Neerhoff and Mur approach,
which uses a Green's function formalism for rigorous numerical solution
of Maxwell's equations for a single, isolated slit \cite{Neer,Betz}.
The transmission of the slit array is determined by calculating all
the light power $P(\lambda)$ radiated into the far-field diffraction
zone, $x{\in}[-\infty,\infty]$ at the distance
$z\gg{\lambda}$ from the screen. The total per-slit transmission
coefficient, which represents the per-slit enhancement in
transmission achieved by taking a single, isolated slit and
placing it in an $M$-slit array, is then found by using an
equation $T_M(\lambda)=P(\lambda)/MP_1$. Here, $P$ is
the total power of $M$ beams produced by the array,
and $P_1$ is the power of a beam produced by the single slit.
Figure 1 shows the transmission
coefficient $T_M(\lambda)$, in the spectral region 500-2000 nm,
calculated for the array parameters: $a=100$ nm, $\Lambda=1800$
nm, and $b=5\times 10^{-3}\lambda_{max}$.
\begin{figure}
\begin{center}
\includegraphics[keepaspectratio, width=1\columnwidth]{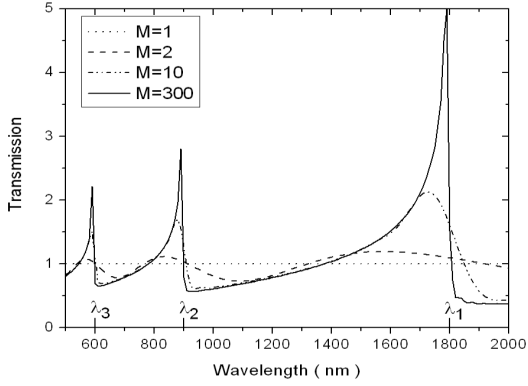}
\end{center}
\caption{The per-slit transmission $T_M(\lambda)$ of an array of
independent slits of the period $\Lambda$ versus the wavelength for different number $M$
of slits. There are three Fabry-Perot like resonances at the
wavelengths $\lambda_n{\approx}\Lambda/n$, $n$=1, 2 and 3.}
\label{fig:Fig1}
\end{figure}
The transmitted power was computed by integrating the total energy
flux at the distance $z$ = 1 mm over the detector region of width
$\Delta{x}$ = 20 mm. The transmission spectra $T_M(\lambda)$ is
shown for different values of $M$. We notice that the spectra
$T_M(\lambda)$ is periodically modulated, as a function of
wavelength, below and above a level defined by the transmission
$T_1(\lambda)=1$ of one isolated slit. As $M$ is increased from 2
to 10, the visibility of the modulation fringes increases
approximately from 0.2 to 0.7. The transmission $T_M$ exhibits the
Fabry-Perot like maxima around wavelengths
$\lambda_{n}=\Lambda/n$ ($n$=1, 2, ...). The spectral peaks
increase with increasing the number of slits and reach a saturation
($T_M^{max}\approx5$) in amplitude by $M=300$, at
$\lambda\approx{1800}$ nm. The peak widths and the spectral shifts
of the resonances from the Fabry-Perot wavelengths decrease with
increasing the number $M$ of slits. Figure~1 indicates that enhancement
and suppression in the transmission spectra are the natural properties of an ensemble of
independent subwavelength slits in a thin ($b\ll\lambda$) screen.
The spectral peaks are characterized by asymmetric Fano-like
profiles. Such modulations in the transmission spectra are known
as Wood's anomalies. The minima and maxima correspond to Rayleigh
anomalies and Fano resonances, respectively. The Wood
anomalies in transmission spectra of
optical gratings, a long standing problem in optics~\cite{Hess}, follows
naturally from interference properties of our model. The new 
point is a weak Wood's anomaly in a classical
Young type two-slit system ($M=2$). Figure 1 shows
the {\em extra energy} ($T > 1$), {\em negative energy} ($T < 1$)
and {\em annihilation of energy} ($T < 1$). The creating and destroying
of energy break the energy conservation law in the system of $M$
independent subwavelength beams (slits).

To clarify the results of the computer code we have developed an
analytical model, which yields simple formulas for the diffracted fields.
For the fields diffracted by a narrow ($2a\ll\lambda, b\geq0$) slit into the
region $|z|> 2a$, it can be shown that the Neerhoff and Mur model
simplifies to an analytical one. For the magnetic
$\vec{H}=(0,H_y,0)$ and electric $\vec{E}=(E_x,0,E_z)$ fields we
found:
\begin{eqnarray}
{H_y}(x,z)=i{a}DF_0^{1}(k[x^2+z^2]^{1/2}),
\end{eqnarray}
\begin{eqnarray}
E_{x}(x,z)={{-az}{[x^2+z^2]^{-1/2}}}D
F_1^{1}(k[x^2+z^2]^{1/2}),
\end{eqnarray}
and
\begin{eqnarray}
E_{z}(x,z)={{ax}{[x^2+z^2]^{-1/2}}}D
F_1^{1}(k[x^2+z^2]^{1/2}),
\end{eqnarray}
where
\begin{eqnarray}
\label{sz:D:def} D=4k^{-1}[[\exp(ikb)(aA-k)]^{2}-(aA+k)^2]^{-1}
\end{eqnarray}
and
\begin{eqnarray}
\label{sz:A:def}
A=F_0^{1}(ka)+\frac{\pi}{2}[\bar{F}_{0}(ka)F_1^{1}(ka)
+\bar{F}_{1}(ka)F_0^{1}(ka)].
\end{eqnarray}
Here, $F_1^{1}$, $F_0^{1}$, $\bar{F}_{0}$ and $\bar{F}_{1}$ are
the Hankel and Struve functions, respectively. The fields are
spatially nonuniform, in contrast to a common opinion that a
subwavelength aperture diffracts light in all directions uniformly
\cite{Lez}. The fields produced by an array of $M$ independent
slits are given by
$\vec{E}(x,z)=\sum_{m=1}^{M}\vec{E}_{m}(x+m\Lambda,z)$ and
$\vec{H}(x,z)=\sum_{m=1}^{M}\vec{H}_{m}(x+m\Lambda,z)$, where
$\vec{E}_{m}$ and $\vec{H}_{m}$ are the fields of an $m$-th beam
generated by the respective slit. As an example, Fig.~2(a)
compares the far-field distributions calculated by the analytical
formulas (1-5) to that obtained in the rigorous model. We notice
that the distributions are indistinguishable. The field power $P$,
which determines the field energy, is found by integrating the
energy flux
$|\langle\vec{S}\rangle_t|=|\vec{E}\times\vec{H}^*+\vec{E}^*\times\vec{H}|(c/16\pi)$.
Thus, the analytical model describes accurately also the energy
balance in the system of $M$ independent subwavelength beams.
\begin{figure}
\begin{center}
\includegraphics[keepaspectratio, width=0.9\columnwidth]{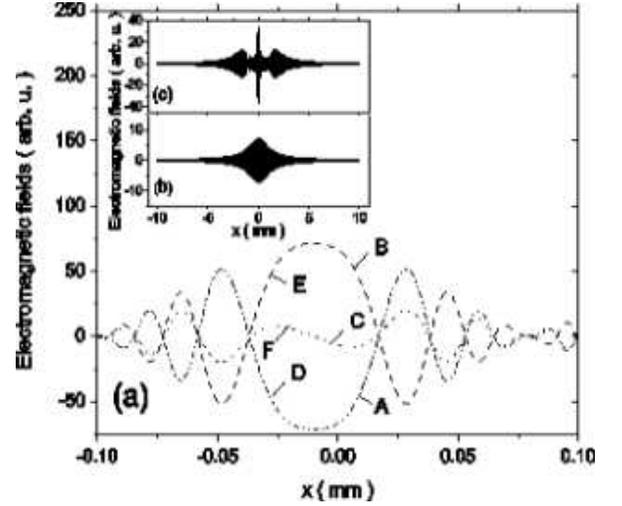}
\end{center}
\caption{Electromagnetic fields in the far-field zone. (a) The
fields Re($E_x(x)$) ($A$ and $D$), Re($H_y(x)$) ($B$ and $E$), and
Re($10E_z(x)$) ($C$ and $F$) calculated for $M$ = 10 and $\lambda$
= 1600~nm. The curves $A$, $B$, and $C$: rigorous model; curves
$D$, $E$, and $F$: analytical model. (b) Re($E_x(x)$) for $M$=1:
analytical model. (c) Re($E_x(x)$) for $M$=5: analytical model.}
\label{fig:Fig2}
\end{figure}
The model does not only support results of our computer code, but
presents an intuitively transparent explanation (physical
mechanism) of the {\em extra energy}, {\em negative energy}, and
{\em annihilation of energy} in terms of the constructive or
destructive interference of the $M$ independent subwavelength
beams produced by the multi-beam source. The creating and
destroying of energy, which are associated with the extraordinary 
transmission without assistance of the surface waves, break the energy 
conservation law. Notice that the array-induced decrease of the central beam
divergence (Figs.~2(b) and 2(c)) is relevant to the beaming light
\cite{Mart}, and the diffraction-free light and matter beams
\cite{Kuk}. The amplitudes of the beams (evanescent fields) can
rapidly decrease with increasing the distances $x$ and $z$.
However, due to the enhancement and beaming mechanisms
(Figs.~1-4), the array produces a propagating wave with low
divergence. Such a behavior is in agreement with the
Huygens-Fresnel principle, which considers a propagating wave as a
superposition of secondary spherical waves.

It is now important to understand the energy balance in the two
fundamental systems of wave optics, the single-slit and two-slit
systems. The major features of the transmission through a single
subwavelength slit are the intraslit resonances and the spectral
shifts of the resonances from the Fabry-Perot wavelengths
\cite{Taka}. In agreement with the predictions \cite{Taka}, the
formula (4) shows that the transmission $T$ = $P/P_0$ =
$(a/k)[$Re$(D)]^{2}+[$Im$(D)]^2$ exhibits Fabry-Perot like maxima
around wavelengths $\lambda_{n}=2b/n$, where $P_0$ is the power
impinging on the slit opening. The enhancement and spectral shifts
are explained by the wavelength dependent terms in the denominator
of Eq. (4). The enhancement ($T(\lambda_1){\approx}b/{\pi}a$
\cite{Kuk}) is in contrast to the attenuation predicted by the
model \cite{Taka}. At the resonant conditions, the system
redistributes the electromagnetic energy in the intra-slit region
and around the screen, such that more energy ($T>1$) is
effectively transmitted compared to the energy impinging on the
slit opening. The total energy of the system is conserved under
the energy redistribution. This mechanism is different from those
based on the creating and destroying of energy by the multi-beam
(multi-slit) system. The Young type two-slit configuration is
characterized by a sinusoidal modulation of the transmission
spectra $T_2(\lambda)$ \cite{Scho,Lalan}. The modulation period is
inversely proportional to the slit separation $\Lambda$. The
visibility $V$ of the fringes is of order 0.2, independently on
the slit separation. In our model, the transmission is given by
$T_{2}{\sim}{\int} [F_1^1(x_1)[iF_0^1(x_2)]^{*}+F_1^1
(x_1)^{*}iF_0^1(x_2)]dx$, where $x_1=x$ and $x_2=x+\Lambda$. The
high-frequency modulations with the sideband-frequency
$f_s(\Lambda)$
${\approx}f_1(\lambda)+{f_2(\Lambda,\lambda)}{\sim}1/{\Lambda}$
(Figs. 1 and 3) are produced like that in a classic heterodyne
system by mixing two waves having different spatial frequencies,
$f_1$ and $f_2$.
\begin{figure}
\begin{center}
\includegraphics[keepaspectratio, width=0.8\columnwidth]{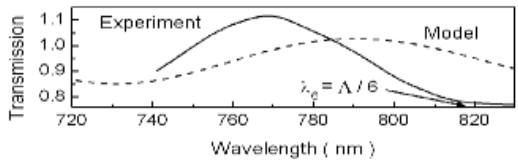}
\end{center}
\caption{The per-slit transmission coefficient $T(\lambda)$ versus
wavelength for the Young type two-slit experiment \cite{Scho}.
Solid curve: experiment; dashed curve: analytical model.
Parameters: $a$ = 100 nm, $\Lambda$ = 4900 nm, and $b$ = 210 nm.}
\label{fig:Fig3}
\end{figure}
Although our model ignores the plasmons, its prediction for the transmission
($T_2^{max}{\approx}1.1$),
the visibility ($V{\approx}0.1$) of the fringes and the resonant
wavelengths $\lambda_{n}\approx\Lambda/n$ compare well with the
plasmon-assisted Young's type experiment \cite{Scho} (Fig.~3). In
the case of $b\geq{\lambda/2}$, the resonances at
$\lambda_{n}=\Lambda/n$ can be accompanied by the intraslit
resonances at $\lambda_{n}=2b/n$. One can easily demonstrate such
behavior by using the analytical formulas (1-5). We considered 
the TM polarization because TE modes are cut off by a thick slit. 
In the case of a thin screen, TE modes propagate into slits so that
magneto-polaritons develop. Because of the symmetry of Maxwell's
equations the scattering intensity is formally identical with
$\vec{E}$ and $\vec{H}$ swapping roles. The described mechanism is not 
the only contribution to enhanced transmission. There 
can be also enhancement due to the energy redistribution by 
surface waves \cite{Schr,Sobn,Port,Leze,Scho,Lalan,Pend}. The surface waves 
can couple the radiation phases of the slits, so that they get 
synchronized, and a collective emission can release the stored 
energy as an enhanced radiation. This kind of enhancement is of 
different nature compared to our model, because the model does 
not require coupling between the beams.

In order to gain physical insight into the energy balance
in the multi-wave ($M{\geq}2$) systems, we
now consider the dependence of the transmission $T_M(\lambda)$ on
the slit separation $\Lambda$. We assume that the slits (beams) are
independent also at $\Lambda\rightarrow{0}$.
\begin{figure}
\begin{center}
\includegraphics[keepaspectratio, width=0.8\columnwidth]{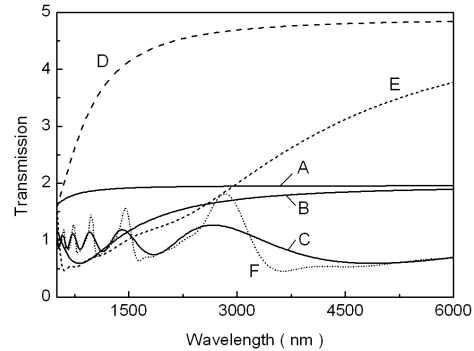}
\end{center}
\caption{The per-slit transmission $T_M(\lambda)$ versus
wavelength for the different values of $\Lambda$ and $M$: (A)
$\Lambda=$ 100 nm, $M=2$; (B) $\Lambda=$ 500 nm, $M=2$; (C)
$\Lambda=$ 3000 nm, $M=2$; (D) $\Lambda=$ 100 nm, $M=5$; (E)
$\Lambda=$ 500 nm, $M=5$; (F) $\Lambda=$ 3000 nm, $M=5$.
Parameters: $a$ = 100 nm and $b$ = 10 nm.  There are two
enhancement regimes at $\Lambda\ll\lambda$ and
$\Lambda{\geq}\lambda$.} \label{fig:Fig3}
\end{figure}
According to the Van Citter-Zernike coherence theorem, a light
source (even incoherent) of radius $r=M(a+{\Lambda})$ produces a
transversally coherent wave at the distance
$z{\geq}{\pi}Rr/\lambda$ in the region of radius $R$. Thus, in the
case of a subwavelength light source ($\Lambda\ll\lambda$), the
collective emission of the ensemble of slits generates the
coherent electric and magnetic fields,
$\vec{E}=\sum_{m=1}^{M}\vec{E}_{m}$exp$(i\varphi_m){\approx}M\vec{E}_{1}$exp$(i\varphi)$
and $\vec{H}{\approx}M\vec{H}_{1}$exp$(i\varphi)$. This means that
the beams arrive at the detector with the same phases
$\varphi_{m}(x)\approx{\varphi{(x)}}$ (see, also Ref.~\cite{Gen}). Consequently,
the power (energy) of the emitted light scales with the square of the number
of slits (beams), $P\approx{M^2}P_1$. Therefore, the transmission
($T_M=P/MP_1$) grows linearly with the number of slits,
$T_M{\approx}M$. For a given $M$, the function $T_M(\lambda)$
monotonically varies with $\lambda$ (Fig. 4). At the appropriate conditions,
the transmission can reach the 1000-times enhancement
($M={\lambda}z/\pi{R}(a+\Lambda)$). In the case
of $R\geq{\lambda}z/{\pi}r$ ($\Lambda{\geq}\lambda$), the beams
arrive at the detector with different phases $\varphi_{m}(x)$.
Consequently, the power and transmission grow more slow with the
number of beams (Figs.~1-4). Notice, that according to the
energy conservation law one should expect that the energy (transmission)
would remain constant with changing the slit (beam) separation.
The creating and destroying of energy in a wave field, which are associated with 
the extraordinary transmission, break the conservation law. In the case 
of ${\Lambda}>>\lambda $, our model is in agreement with the conservation 
law and theories of gratings~\cite{Pet}. Our consideration of the subwavelength gratings
is similar in spirit to the dynamical diffraction models \cite{Trea,Gar},
the Airy-like model \cite{Cao}, and especially to a surface evanescent
wave model~\cite{Leze}.

One can easily demonstrate the breaking of energy conservation in
any subwavelength physical system by taking into account the
interference properties of Young's double-source system. At the
risk of belaboring the obvious, we now describe the phenomenon.
In the far-field diffraction zone, the radiation from two pinholes
of Young's setup is described by two spherical waves. The light
intensity at the detector is $I(\vec
r)=|(E/r_1)exp(ikr_1+\varphi_1)+
(E/r_2)exp(ikr_2+\varphi_2)|^2=I_1+I_2+2(I_1I_2)^{1/2}cos([kr_1+\varphi_1]-[kr_2+\varphi_2])$.
The corresponding energy is
$W=\int{\int}(I_1+I_2+2(I_1I_2)^{1/2}cos([kr_1+\varphi_1]-[kr_2+\varphi_2])dxdy$.
Here, we use the units $c{\Delta}t/8\pi=1$. In conventional
Young's setup, which contains the pinholes separated by the
distance ${\Lambda}>>\lambda $, the interference cross term
(energy) vanishes. In accordance with the conservation law, the
energy is $W=\int{\int}(I_1+I_2)dxdy=W_1+W_2=2W_0$,  where
$W_1=W_2=W_0$. In the case of Young's subwavelength system
(${\Lambda}<<\lambda$; $r_1=r_2$ for any coordinate $x$ or $y$),
the energy is
$W=W_1+W_2+2\int{\int}(I_1I_2)^{1/2}cos(\varphi_1-\varphi_2)dxdy$.
The first-order correlation term is the {\em positive} or {\em negative} 
{\em extra} {\em energy}. 
At the phase condition $\varphi_1-\varphi_2=\pi$, the
interference completely {\em destroys} ($W=0$) the energy. The system
{\em creates} energy ($W=4W_0$) in the case of $\varphi_1-\varphi_2=0$.
The same phase conditions provide the creating or destroying of
energy by quantum two-source interference (for example, see
formulas 4.A.1-4.A.9 \cite{Scul}). The phenomenon depends neither
on the nature (light or matter) of the waves (continuous waves or
pulses) nor on material and shape of the subwavelength apertures
(1-D and 2-D apertures or fibres). There is an evident resemblance
between our model and a Dicke superradiance quantum model
\cite{Dick} of emission of an ensemble of atoms. A quantum
reformulation of our model, which will be presented in our next
paper, help us to understand better why a quantum entangled state
is preserved on passage through a hole array \cite{Alte}.

It should be stressed that energy in conventional physical systems
can be converted from one form to another, but it cannot be created
or destroyed. According to our model, energy may be created or
destroyed by constructive or destructive interference of waves (beams)
only at the extremely particular phase conditions. The interference
completely destroys energy if waves interfere destructively in
all points of a physical system. The interference creates energy if
waves interfere only constructively. The experimental realization
of such phase conditions is practically impossible in conventional
physical systems. We showed that the waves generated by the point-like
sources separated by the distance ${\Lambda}<\lambda $ (for example,
subwavelength gratings) can satisfy the phase conditions in the far-field
diffraction zone.

This study was supported by the Hungarian Scientific
Research Foundation (OTKA, Contract No T046811).

\end{document}